\documentclass[12pt]{article}
\usepackage{graphics}
\makeatletter

\renewcommand{\section}{\@startsection {section}{1}{\z@}
{-3.5ex plus -1ex minus -.2ex}{2.3ex plus .2ex}{\normalsize\bf}}

\renewcommand{\subsection}{\@startsection{subsection}{2}{\z@}
{-3.25ex plus -1ex minus -.2ex}{1.5ex plus .2ex}{\normalsize\it}}

\def\abstract{\if@twocolumn
\section*{\abstractname}
\else \small
\quotation
\fi}
\def\endabstract{\if@twocolumn\else\endquotation\fi}

\renewcommand{\@makefnmark}{\hbox{\mathsurround=0pt
$^\dagger$}}
\renewcommand{\@makefntext}[1]{\parindent=1em\noindent
\hbox to 1.8em{\hss$^\dagger$}#1}

\def\thebibliography#1{\section*{\refname\@mkboth
 {\uppercase{\refname}}{\uppercase{\refname}}}\list
 {[\arabic{enumi}]}{\settowidth\labelwidth{[#1]}\leftmargin\labelwidth
 \advance\leftmargin\labelsep\parsep=0em\itemsep=0em
 \usecounter{enumi}}
 \def\newblock{\hskip .11em plus .33em minus .07em}
 \sloppy\clubpenalty4000\widowpenalty4000
 \sfcode`\.=1000\relax}

\makeatother

\begin{document}

\begin{center}

{\normalsize\bf ON THE COMPETITION BETWEEN FERROMAGNETIC AND
ANTIFERROMAGNETIC STATES IN Sr$_2$MnMoO$_6$
}\\
\bigskip
I.~V.~Solovyev\footnote{Corresponding author. Fax: +81 (298) 61
2586; e-mail: igor.solovyev@aist.go.jp}
\medskip \\
{\small\it
Tokura Spin Superstructure Project (SSS), \\
ERATO Japan Science and Technology Corporation, \\
c/o National Institute of Advanced Industrial Science and Technology, \\
AIST Tsukuba Central 4, 1-1-1 Higashi, Tsukuba, Ibaraki 305-8562,
Japan}

\end{center}

\begin{abstract}

It is argued that the magnetic behavior of Sr$_2$MnMoO$_6$ is
determined by the existence of two total energy minima
corresponding to the metallic ferromagnetic and insulating
antiferromagnetic states, which may be nearly degenerate depending
on the magnitude of the breathing distortion.


\end{abstract}

\section{Introduction}

  Recently, the ordered double perovskites Sr$_2$$M$$M'$O$_6$ ($M$ being
the magnetic $3d$ element, and $M'$ being the $4d$ or $5d$
element) have attracted a great deal of attention. The interest to
these systems was spurred on by large intergrain-tunneling
magnetoresistance observed at room temperature in Sr$_2$FeMoO$_6$
and Sr$_2$FeReO$_6$~\cite{Kobayashi}, which opens wide
perspectives for technological applications. On the contrary,
Sr$_2$MnMoO$_6$ remains paramagnetic down to the very low
temperature~\cite{Moritomo}. The magnetic susceptibility of
Sr$_2$MnMoO$_6$ obeys the Curie-Weiss law with fairly large
effective moment. However, there is something that prevents
formation of the long-range magnetic order in this material (a
similar situation occurs in double-layer manganites
La$_{2-2x}$Sr$_{1+2x}$Mn$_2$O$_7$ around $x$$=$$0.7$~\cite{Ling}).
In this work we argue that such a behavior may be due to the
competition between ferromagnetic (FM) and type-II
antiferromagnetic (AFM) phases.

\section{Total energy and local stability of the collinear state}

  The main idea of this work is to complement the standard total energy calculations
with the analysis of inter-atomic magnetic interactions in the
collinear magnetic state characterized by the directions $\{ {\bf
e}^0_{\bf i} \}$ of the spin magnetic moments. The latter can be
described in terms of a perturbation theory expansion for small
rotations $\{ \mbox{\boldmath$\delta\varphi$}_{\bf i} \}$ near $\{
{\bf e}^0_{\bf i} \}$. The ''perturbed'' direction of the magnetic
moment at the site ${\bf i}$ is given by ${\bf e}_{\bf i}$$=$${\bf
e}^0_{\bf i}$$+$$ [\mbox{\boldmath$\delta\varphi$}_{\bf
i}$$\times$${\bf e}^0_{\bf i}]$$
-\frac{1}{2}(\mbox{\boldmath$\delta\varphi$}_{\bf i})^2{\bf
e}^0_{\bf i}$, and the total energy change $\Delta E$$=$$E(\{ {\bf
e_i} \})$$-$$E(\{ {\bf e}^0_{\bf i} \})$ in the second order of
$\{ \mbox{\boldmath$\delta\varphi$}_{\bf i} \}$ can be exactly
mapped onto the Heisenberg model as~\cite{Liechtenstein}:
\begin{equation}
\Delta E = -\frac{1}{2} \sum_{\bf im} J_{\bf m} \left[ {\bf
e}_{\bf i} \cdot {\bf e}_{{\bf i}+{\bf m}} - {\bf e}^0_{\bf i}
\cdot {\bf e}^0_{{\bf i}+{\bf m}} \right]. \label{eqn:Heisenberg}
\end{equation}
Eq.~(\ref{eqn:Heisenberg}) originates from the Taylor series
expansion for $E(\{ {\bf e_i} \})$ near $\{ {\bf e}^0_{\bf i} \}$,
and $\{ J_{\bf m} \}$ are related with the second derivatives of
$E(\{ {\bf e_i} \})$ with respect to $\{
\mbox{\boldmath$\delta\varphi$}_{\bf i} \}$. $\{ J_{\bf m} \}$ can
be expressed through the matrix elements of the one-electron Green
function and the magnetic part of the Kohn-Sham potential at the
site ${\bf m}$, $\Delta_{\rm ex}^{\bf m}$$=$$\frac{1}{2}(v_{\bf
m}^\uparrow$$-$$v_{\bf m}^\downarrow)$, as~\cite{Liechtenstein}:
\begin{equation}
J_{\bf m}=\frac{1}{2\pi} {\rm Im} \int_{-\infty}^{\varepsilon_F} d
\varepsilon {\rm Tr}_L \left\{ \Delta^{\bf 0}_{\rm ex}
\widehat{\cal G}^\uparrow_{\bf 0m}(\varepsilon) \Delta^{\bf
m}_{\rm ex} \widehat{\cal G}^\downarrow_{\bf m0}(\varepsilon)
\right\}, \label{eqn:Jm}
\end{equation}
with ${\rm Tr}_L$ denoting the trace over the orbital indices.

  All calculations have been performed in the local-spin-density approximation,
using the ASA-LMTO method~\cite{LMTO}.

\section{Electronic structure and magnetic interactions}

\begin{figure}[ht]
\centering \noindent
\resizebox{7cm}{!}{\includegraphics{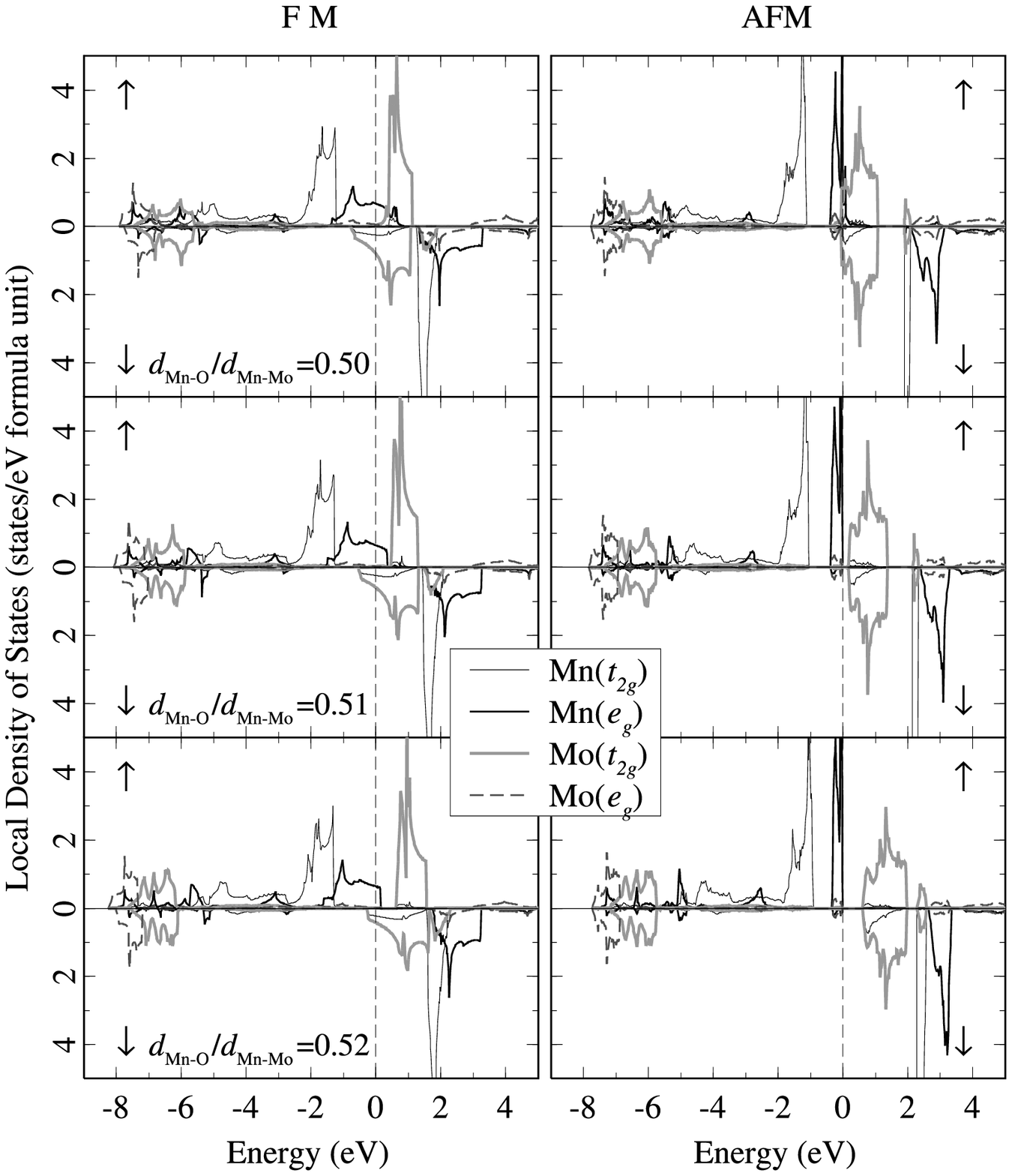}}
\caption{Local
densities of states for the ferromagnetic (left) and type-II
antiferromagnetic (right) phases of Sr$_2$MnMoO$_6$ as a function
of breathing distortion.} \label{fig.DOS}
\end{figure}

  The calculated densities of states for the FM and
AFM phases are shown in Fig.\ref{fig.DOS}, as a function of the
breathing distortion $\delta$$=$$d_{\rm Mn-O}/d_{\rm Mn-Mo}$ (the
ratio of Mn-O and Mn-Mo bondlengths).

  When $\delta$$\leq$$0.52$, the FM state is metallic.
The Fermi level crosses Mn($e_g$) band in the majority
($\uparrow$)-spin channel and the Mo($t_{2g}$) band in the
minority ($\downarrow$)-spin channel. Therefore, one can expect
two main contributions, which tends to stabilize the FM ordering:
the canonical double exchange (DE) operating in the
$\uparrow$-spin Mn($e_g$) band, similar to the colossal
magnetoresistive manganites~\cite{Springer02}; and a generalized
DE mechanism associated with partial filling of the
$\downarrow$-spin Mo($t_{2g}$) band and originating from the
strong Mn-Mo hybridization~\cite{Fang,comment1}. The breathing
distortion $\delta$$>$$0.5$ tends to fully populate the Mn($e_g$)
band and depopulate the Mo($t_{2g}$) band. Therefore, when
$\delta$ increases, the DE interactions of the both types will
{\it decreases}.

  The AFM phase is metallic in the undistorted cubic structure
($\delta$$=$$0.5$). However, even small oxygen displacement
towards the Mo sites opens the band gap. The situation corresponds
to the formal configurations $3d^5_\uparrow 3d^0_\downarrow$
(Mn$^{2+}$) and $4d^0$ (Mo$^{6+}$) of the transition-metal sites.
The half-filled Mn($3d$) states give rise to the AFM superexchange
(SE) interactions, similar to the rock-salt monoxide MnO, which
are roughly proportional to $-$$t_{\rm eff}^2/\Delta_{\rm ex}^{\rm
Mn}$. The effective hoppings $t_{\rm eff}$ between
nearest-neighbor ($nn$) Mn($3d$) orbitals are smaller in
Sr$_2$MnMoO$_6$ (in comparison with MnO), as they are mediated by
much longer Mn-O-Mo-O-Mn paths. However, $\Delta_{\rm ex}^{\rm
Mn}$ is expected to be also smaller in Sr$_2$MnMoO$_6$, which is
more itinerant material than MnO. Thus, smaller $t_{\rm eff}$ in
Sr$_2$MnMoO$_6$ is compensated by smaller $\Delta_{\rm ex}^{\rm
Mn}$, and the resulting SE interactions are expected to be
comparable with those in MnO.

\begin{figure}[ht]
\centering \noindent
\resizebox{7.0cm}{!}{\includegraphics{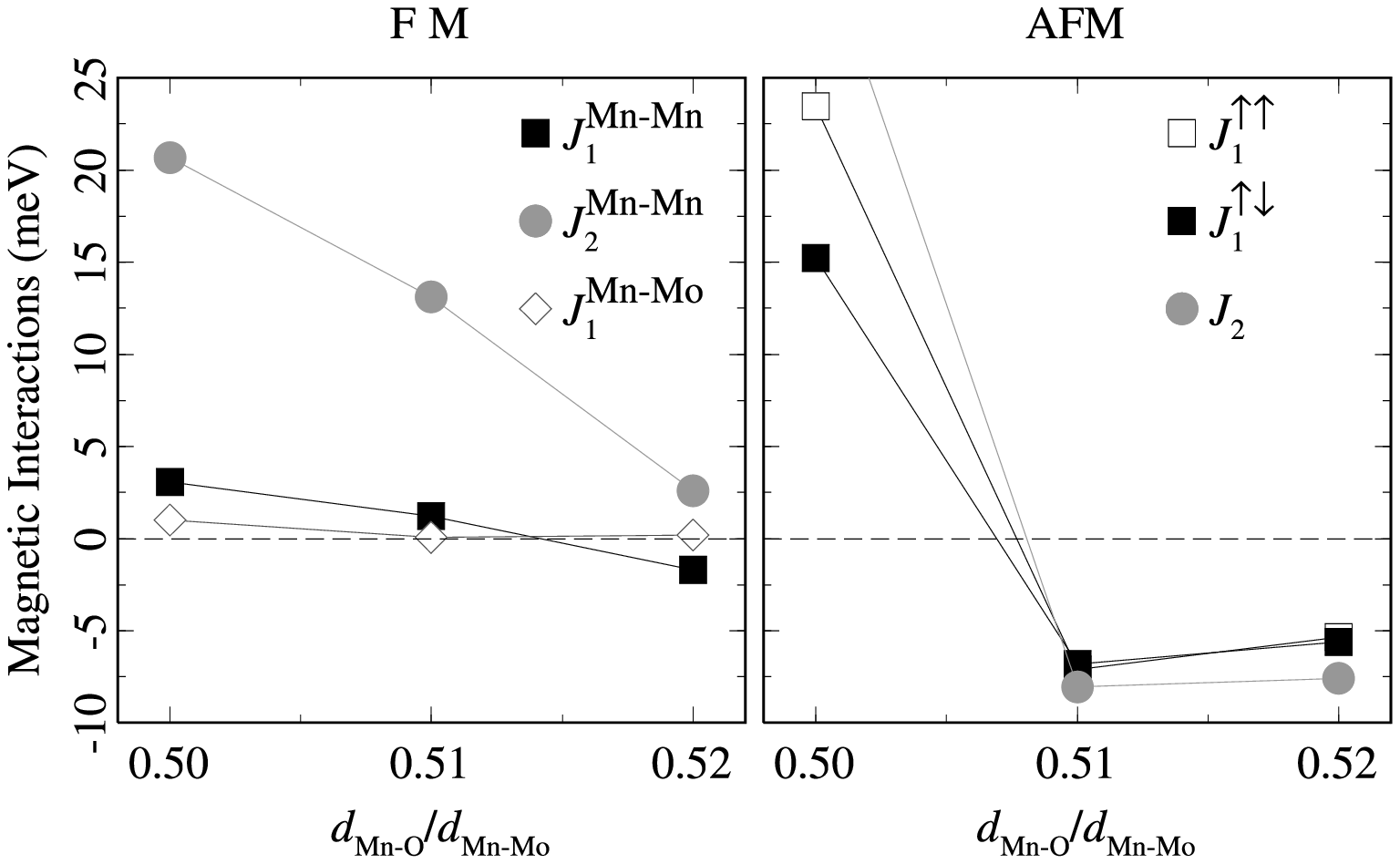}}
\caption{Inter-atomic magnetic interactions calculated in the
ferromagnetic (left) and antiferromagnetic (right) states of
Sr$_2$MnMoO$_6$ as a function of breathing distortion.
$J_1^{\uparrow \uparrow}$ and $J_1^{\uparrow \downarrow}$ are the
nearest-neighbor interactions between Mn sites with the same and
opposite directions of spins in the type-II antiferromagnetic
structure.} \label{fig.Jij}
\end{figure}

  The qualitative discussions are supported by direct calculations of inter-atomic magnetic
interactions (Fig.\ref{fig.Jij}). In the FM state, the $180^\circ$
exchange $J_2^{\rm Mn-Mn}$ between next-nearest-neighbors in the
Mn sublattice is the strongest interaction. The $nn$ ($90^\circ$)
interaction $J_1^{\rm Mn-Mn}$ is significantly weaker and becomes
antiferromagnetic around $\delta$$=$$0.51$. However, the FM
interaction $J_2^{\rm Mn-Mn}$ largely prevails and the FM phase
remains stable at least up to $\delta$$=$$0.52$. The $nn$
interactions $J_1^{\rm Mn-Mo}$ between Mn and Mo sublattices is
small and does not play any role in the problem.

  The breathing distortion in the AFM state stabilizes the AFM
coupling (both between first and second nearest neighbors in the
Mn-sublattice). $J_1$ and $J_2$ are of the order of 5-8 meV, and
comparable with similar interactions in MnO~\cite{PRB98}.

\begin{figure}[ht]
\centering \noindent
\resizebox{6.0cm}{!}{\includegraphics{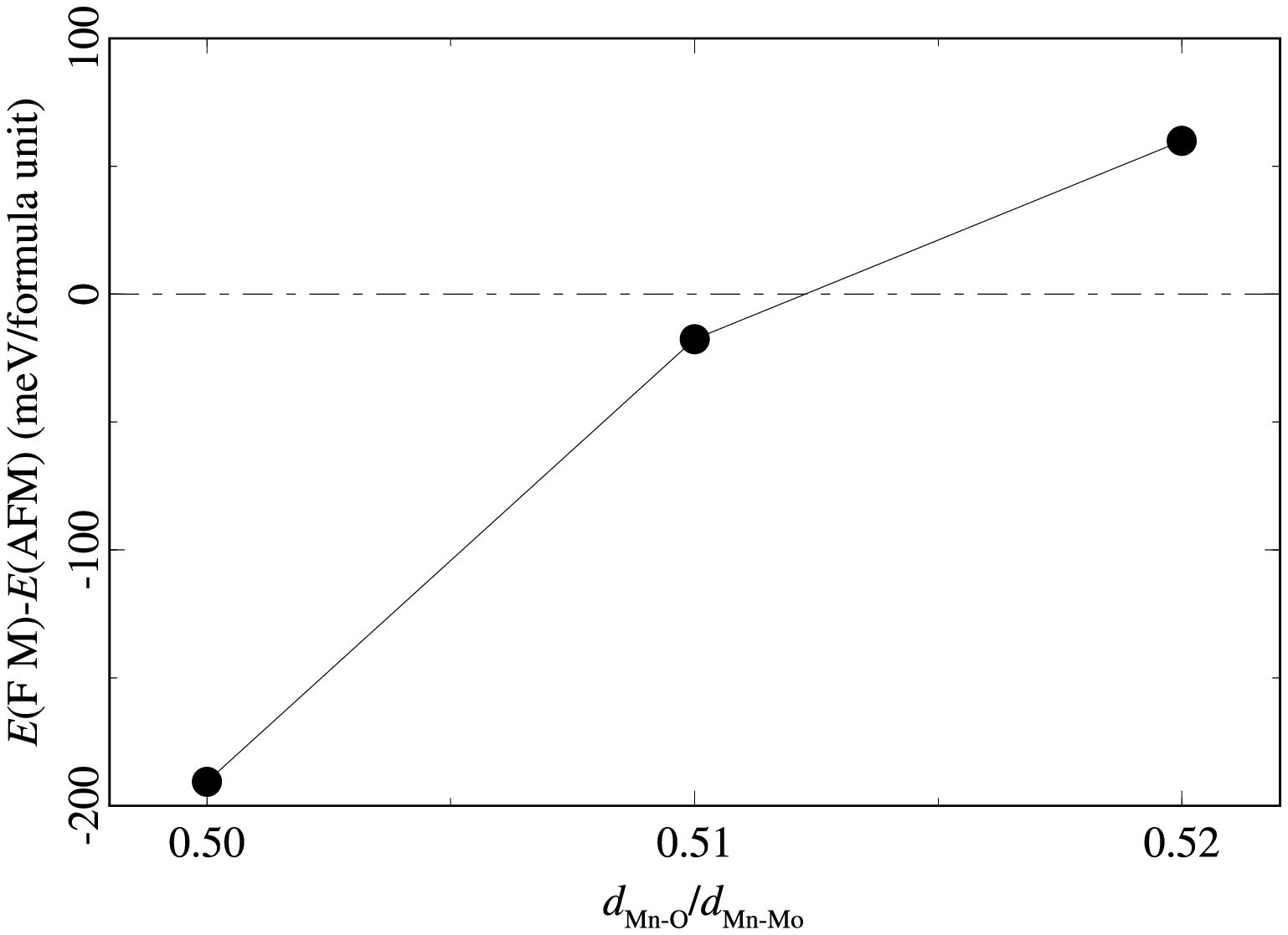}} \caption{Total
energy difference between ferromagnetic and antiferromagnetic
configurations as a function of breathing distortion.}
\label{fig.etot}
\end{figure}

  Thus, within the interval
$0.5$$<$$\delta$$\leq$$0.52$ {\it both} FM and type-II AFM phases
are locally stable. Next, we argue that these two local minima of
the total energy may exist in the very narrow energy range, so
that {\it two magnetic solutions are nearly degenerate}. This is
qualitatively supported by results of direct total energy
calculations shown in Fig.\ref{fig.etot}. The situation when the
FM and AFM solutions become nearly degenerate can occur roughly
around $\delta$$=$$0.51$.

\section{Conclusions}

  We have argued that Sr$_2$MnMoO$_6$ may have
(at least) two total energy minima corresponding to the metallic
FM and insulating type-II AFM states, which coexist in the very
narrow energy range. We expect the competition between these
states to prevent the formation of the long-range magnetic order
which features the experimental behavior of Sr$_2$MnMoO$_6$ down
to the very low temperature. The quantitative description of such
a situation, using results of first-principles electronic
structure calculations, could be one of the possible extensions of
the present work.

\small{

}  

\begin{thebibliography}{10}
\bibitem{comment1}
Since there is no direct hoppings between the
$e_g$ and $t_{2g}$ orbitals in the cubic lattice, these two
contributions can be considered separately.
\end{thebibliography}
\end{document}